# Petahertz-bandwidth spectral combs from the amplified spontaneous emission of a collisional plasma-based x-ray laser

I. R. Khairulin[1,2], V. A. Antonov[1], I. D. Morskov[2], M. Yu. Ryabikin[1,2]

[1] *Gaponov-Grekhov Institute of Applied Physics of the Russian Academy of Sciences,
46 Ulyanov Street, Nizhny Novgorod 603950, Russia*
[2] *Lobachevsky State University of Nizhny Novgorod,
23 Gagarin Avenue, Nizhny Novgorod 603950, Russia*

Corresponding author: I. R. Khairulin, khairulinir@ipfran.ru

The possibility of generating spectral combs of petahertz bandwidth is demonstrated from amplified spontaneous emission (ASE) of a collisional plasma-based x-ray laser when its active medium is irradiated with a linearly polarized near-infrared (IR) laser field. It is shown that at a sufficiently high IR-field intensity these combs correspond to a sequence of subfemtosecond beats of extreme ultraviolet (XUV) radiation of a deterministic shape. At the same time, at a lower IR-field intensity ASE remains quasimonochromatic, but its frequency shifts as a result of the IR-field-induced quadratic Stark effect. In this case, the frequency tuning range of XUV radiation is approximately twice the frequency of the modulating IR field. In addition, modulation by the IR field leads to amplitude and frequency discrimination of the polarization components of ASE, parallel and orthogonal to the IR field. The possibility of observing these effects is analyzed for a collisional plasma x-ray laser based on neon-like $Ti^{12+}$ ions with an unperturbed ASE wavelength of 32.6 nm and a modulating field with a wavelength of 1 μm and an intensity in the range of $10^{16} - 10^{17}$ W/cm².

## I. INTRODUCTION

The basis of modern scientific and technological progress is largely the use of short-wave electromagnetic radiation of the extreme ultraviolet (XUV) and x-ray ranges. In recent decades, sources of such radiation have been proposed based on various physical principles.

One of the main classes of such sources are plasma x-ray lasers, based on the creation of a population inversion at the transition (usually between excited states) of multiply charged ions during the evolution of plasma created due to ionization of a solid target by a sequence of laser pulses in the near infrared (IR) range [1-3]. A modern plasma-based x-ray laser pumped by a traveling wave of an IR field is a source of unidirectional amplified spontaneous emission (ASE) in the XUV/x-ray range [4, 5]. Its active medium has the shape close to a thin elongated cylinder, and its optical thickness (the product of the gain coefficient for resonant XUV/x-ray radiation by the physical thickness of the medium) can reach several dozen. This allows generating directed radiation with a high degree of temporal coherence (the typical ratio of the spectral width to the carrier frequency is about $10^{-5}$) and a sufficiently high energy, varying from μJ for radiation with a wavelength of about 10–20 nm to several mJ for radiation with a wavelength of about 50 nm. To date, the minimum achieved wavelength of optically pumped soft x-ray laser generation is 4.03 nm [6], while gain saturation has been demonstrated for a wavelength of 6.85 nm [7]. Plasma x-ray laser radiation is actively used for high resolution microscopy, micro-holography, very high plasma density measurements, applications to semiconductor surface studies, and nanolithography [8-13]. At the same time, the lack of frequency tuning capability of plasma-based x-ray lasers limits their spectroscopic applications, while the picosecond duration of the generated pulses prevents their use for studying and controlling physical processes occurring on femto- and attosecond time scales.

It should be noted that the duration of the radiation pulses of plasma-based x-ray lasers can be reduced to hundreds of femtoseconds by using high harmonics of the optical field as a seed [14-18], which also allows for an additional increase in the spatial coherence of radiation. However, a further reduction in the duration of the generated pulses is impossible due to the limited width of the gain spectrum, which is determined by the parameters of the plasma active medium of the x-ray laser. In this case, out of a large number of harmonics in the spectrum of the seed, only one harmonic is amplified, the frequency of which coincides with the frequency of the inverted transition.

To overcome the limitation associated with the small width of the generation and amplification spectrum of a plasma-based x-ray laser, recent studies [19–27] have proposed irradiating its active medium with an intense modulating IR field, the pulse of which should follow with a certain delay relative to the pump pulses, sufficient to form a population inversion by its arrival. The role of the modulating field is to induce a quasistatic Stark shift of the energy levels of the active medium, the constant component of which causes a shift in the central frequency of the inverted transition, while the time-dependent component leads to an enrichment of the gain spectrum with combination frequencies separated from each other by the frequency or the doubled frequency of the modulating field, depending on the type of the ions. It has been theoretically shown that this approach allows (i) to convert the radiation of a single high-order harmonic into a sequence of subfemtosecond pulses with a carrier frequency in XUV or x-ray range [19, 21, 22], (ii) to amplify a sequence of attosecond pulses formed by a set of high harmonics of the modulating IR field [20, 23, 25], and (iii) to control the polarization of high harmonics during their amplification [24, 26, 27]. However, all the above-mentioned studies assumed that at the input of the active medium of the x-ray laser there is a seed radiation, the pulse of which overlaps in time and space with the pulse of the modulating field.

In this work, we investigate the spatial, temporal, spectral, and polarization properties of the ASE directly generated by an optically modulated active medium of a collisional plasma-based x-ray laser in the absence of seed radiation. The plasma of neon-like $Ti^{12+}$ ions [28-31] is considered as the active medium. The paper is organized as follows. The Introduction (Sec. I) is followed by basic information about the active medium of neon-like $Ti^{12+}$ ions and the equations used to describe the ASE generation in the presence of a modulating optical field (Sec. II). Section III presents an analytical solution for the spectral power density of the optically modulated plasma-based x-ray laser and formulates the conclusions that follow from it. In Sec. IV, the analytical solution is compared with the results of the numerical solution of the Maxwell-von Neumann equations. On this basis, the two optical modulation regimes are analyzed, corresponding to the generation of (a) frequency-tunable quasimonochromatic XUV radiation and (b) frequency combs forming a sequence of subfemtosecond pulses (beats). This is followed by the Conclusion (Sec. V) and Appendix containing the derivation of the analytical solution presented in Sec. III.

## II. THEORETICAL MODEL

The description of the generation of ASE of optically modulated active plasma of neon-like $Ti^{12+}$ ions presented below is based on the theoretical model outlined in detail and used in [24, 26, 27] to describe the amplification of high-order harmonics. Accordingly, the main expressions of this section are similar to those of the above articles.

Let us consider the active medium of a plasma x-ray laser based on neon-like $Ti^{12+}$ ions. In it, population inversion is reached at transition $|3p^1S_0\rangle \leftrightarrow |3p^1P_1\rangle$ with an unperturbed wavelength of 32.6 nm between two excited energy levels of $Ti^{12+}$ ions [28-31] (see Fig. 1). The upper energy level is non-degenerate and corresponds to state $|1\rangle = |3p^1S_0, J=0, M=0\rangle$, where $J$ and $M$ are the total orbital angular momentum of the state and its projection onto the quantization axis. In turn, the lower energy level is triply degenerate and corresponds to states $|2\rangle = |3p^1P_1, J=1, M=0\rangle$, $|3\rangle = |3p^1P_1, J=1, M=1\rangle$, and $|4\rangle = |3p^1P_1, J=1, M=-1\rangle$. In an experiment, such a medium can be created by irradiating a polished Ti sample with two laser pulses following each other with a nanosecond delay. The first of these is a nanosecond or subnanosecond laser pulse with an energy of hundreds of millijoules to several joules. It is directed at the sample perpendicular to the surface and focused into a strip ~50 μm wide and up to 1 cm long, resulting in the creation of a plasma of neon-like ions on the sample surface. Next, a second laser pulse with an energy comparable to the energy of the first pulse, but with a significantly shorter duration (from one to several picoseconds), is directed into this plasma at a grazing angle [4, 5, 30, 31]. This pulse induces the creation of a population inversion at the lasing transition of neon-like ions as a result of its absorption and subsequent plasma evolution. It should be noted that during the plasma evolution the distribution of the population inversion on the lasing transition changes both in time and space and depends significantly on the electron concentration. The most suitable for generating and amplifying resonant radiation is the peripheral region of the plasma with an electron concentration of ~0.7–1.5×$10^{20}$ cm$^{-3}$, where the lifetime of the population inversion on the lasing transition is ~10–20 ps, and the local value of the intensity gain is ~40–70 cm$^{-1}$ (two times less in amplitude) (Fig. 1).

Below, for simplicity, we will assume that the active plasma region in which the ASE generation is considered is homogeneous and has the shape of a cylinder elongated along the $x$ axis with thickness $L$ and cross-sectional radius $R \ll L$ (in the calculation results presented below, $R = 10$ μm is implied), see Fig. 1. As in [24, 26, 27], we will assume that the electron concentration is $N_e = 5\times10^{19}$ cm$^{-3}$ and the $Ti^{12+}$ ion concentration is $N_{ion} = 4.2\times10^{18}$ cm$^{-3}$. In this case, the unperturbed gain coefficient for intensity at the lasing transition is 70 cm$^{-1}$ and that for amplitude is $g_0 = 35$ cm$^{-1}$. Note that due to the rapid depletion of states $|2\rangle - |4\rangle$ due to radiative transitions to the lower energy levels of the $Ti^{12+}$ ion, their populations by the time of the onset of ASE generation, which is further taken as the initial moment of time, are close to zero. In this case, approximately 1% of all $Ti^{12+}$ ions are in the upper state $|1\rangle$ at the initial moment of time.

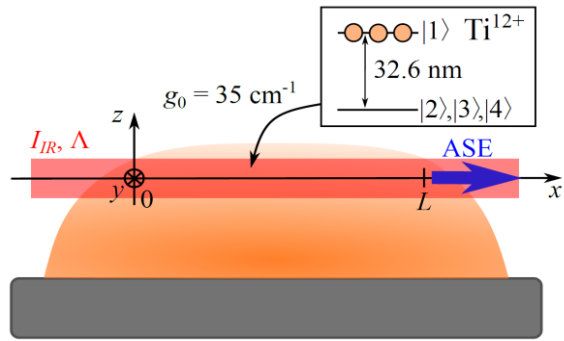

FIG. 1. Schematic diagram of the experiment on ASE generation by optically modulated active medium of plasma x-ray laser based on neon-like $Ti^{12+}$ ions. Modulating optical field with intensity

$I_{IR}$ and frequency $\Omega$ is introduced into the peripheral region of preliminarily created active plasma of neon-like $Ti^{12+}$ ions, in which the characteristic value of unperturbed gain (by amplitude) at the resonance transition $g_0 = 35$ cm$^{-1}$, and the concentration of free electrons is $N_e = 5\times10^{19}$ cm$^{-3}$. The irradiated plasma region is close in shape to a cylinder elongated along the $x$ axis with cross-sectional radius $R$ and length $L \gg R$. In the calculations given below, it is assumed that $R = 10$ μm and $L \leq 1$ cm.

Next, we will assume that the generation region is irradiated by laser radiation of the IR range with a frequency $\Omega$ and linear polarization along the $z$ axis, which will be chosen as the quantization axis, see Fig. 1. We will assume that the duration of the IR field pulse significantly exceeds the characteristic time scales of the processes considered below, and its amplitude changes insignificantly during propagation in the active medium. In this case, the IR field can be represented as a plane monochromatic wave, the electric field of which has the form

$$\vec{E}_{IR}(x,t) = \vec{z}_0 E_{IR} \cos\left[\Omega\left(t - x\sqrt{\varepsilon_{pl}^{(IR)}}/c\right)\right], \tag{1}$$

where $E_{IR}$ is the amplitude of the electric field of the laser wave, $\vec{z}_0$ is the unit polarization vector, $\varepsilon_{pl}^{(IR)} = 1 - \omega_{pl}^2/\Omega^2$ is the permittivity of the plasma at the frequency of the IR field, $\omega_{pl} = \sqrt{4\pi N_e e^2/m_e}$ is the plasma frequency, $e$ and $m_e$ are the charge and mass of the electron, respectively, and $c$ is the speed of light in a vacuum. Next, we will consider such values of $\Omega$ and $E_{IR}$, at which the frequency of the IR field and the Rabi frequencies at all dipole-allowed transitions from states $|1\rangle$, $|2\rangle$, $|3\rangle$, and $|4\rangle$ of the $Ti^{12+}$ ion are much lower than the frequencies of these transitions. In this case, the influence of field (1) is mainly reduced to an instantaneous change in the energies of states in time and space on the scale of the period and wavelength of the laser field due to the quadratic Stark effect [32], and the corresponding transition frequencies are determined as follows:

$$\omega_{12}(t,x) = \bar{\omega}_z + \Delta_z \cos\left[2\Omega\left(t - x\sqrt{\varepsilon_{pl}^{(IR)}}/c\right)\right],$$
$$\omega_{13}(t,x) = \omega_{14}(t,x) = \bar{\omega}_y + \Delta_y \cos\left[2\Omega\left(t - x\sqrt{\varepsilon_{pl}^{(IR)}}/c\right)\right], \tag{2}$$
$$\omega_{23}(t,x) = \omega_{24}(t,x) = (\Delta_z - \Delta_y)\left\{1 + \cos\left[2\Omega\left(t - x\sqrt{\varepsilon_{pl}^{(IR)}}/c\right)\right]\right\},$$
$$\omega_{34}(t,x) = 0,$$

where $\Delta_z \equiv \Delta_{12}$, $\Delta_y \equiv \Delta_{13} = \Delta_{14}$, $\Delta_{ij} = \sum_{k\neq i}(|d_{ki}^{(z)}|E_{IR})^2/(2\hbar^2\omega_{ik}^{(0)}) - \sum_{k\neq j}(|d_{kj}^{(z)}|E_{IR})^2/(2\hbar^2\omega_{jk}^{(0)})$ is the amplitude of the modulation of the $|i\rangle \leftrightarrow |j\rangle$ transition frequency, $d_{ki}^{(z)}$ and $\omega_{ik}^{(0)}$ are the dipole moment and unperturbed frequency of the transition from state $|i\rangle$ to state $|k\rangle$, $\hbar$ is the reduced Planck constant, and the summation is carried out over all states of the $Ti^{12+}$ ion in the absence of the field (1). In addition, the following notations are used in (2): $\bar{\omega}_z = \omega_{12}^{(0)} + \Delta_z$ is the time-average $|1\rangle \leftrightarrow |2\rangle$ transition frequency and $\bar{\omega}_y = \omega_{13}^{(0)} + \Delta_y = \omega_{14}^{(0)} + \Delta_y$ is the time-average $|1\rangle \leftrightarrow |3\rangle,|4\rangle$ transition frequency. Note that for neon-like $Ti^{12+}$ ions, the modulation amplitudes of the transitions $|1\rangle \leftrightarrow |2\rangle$ and $|1\rangle \leftrightarrow |3\rangle,|4\rangle$ differ slightly from each other, $|\Delta_z|/|\Delta_y| \approx 0.93$ (in atomic

units $\Delta_z/E_{IR}^2 \approx 0.1244$ a.u. and $\Delta_y/E_{IR}^2 \approx 0.1343$ a.u.). As a result, the time-averaged frequencies of the above transitions also turn out to be different.

Since at the initial moment of time there is a population inversion at the transitions $|1\rangle \leftrightarrow |2\rangle, |3\rangle, |4\rangle$, then as a result of spontaneous transitions in the active medium, spontaneous radiation will be generated with a carrier frequency $\omega_{XUV}$ in the vicinity of the $|1\rangle \leftrightarrow |2\rangle, |3\rangle, |4\rangle$ transition frequencies. As a result of the amplification of this radiation during propagation along the plasma channel, ASE will appear at its output. Since the characteristic transverse size of the active plasma irradiated by the laser field (1) is significantly smaller than the length of the plasma channel, $R \ll L$, the generation and propagation of ASE can be described using a one-dimensional wave equation, neglecting the change in the characteristics of the generated radiation in the transverse direction:

$$\frac{\partial^2 \vec{E}}{\partial x^2} - \frac{\varepsilon_{pl}^{(XUV)}}{c^2} \frac{\partial^2 \vec{E}}{\partial t^2} = \frac{4\pi}{c} \frac{\partial^2 \vec{P}}{\partial t^2}, \qquad (3)$$

where $\vec{E}$ is the vector of the electric field of the ASE generated in the medium, $\varepsilon_{pl}^{(XUV)} = 1 - \omega_{pl}^2/\omega_{XUV}^2$ is the permittivity of the plasma at frequency $\omega_{XUV}$, and $\vec{P}$ is the vector of the resonant polarization of the medium excited by the generated ASE and defined as

$$\vec{P}(x,t) = N_{ion}^{(res)} \left( \rho_{12} \vec{d}_{21} + \rho_{13} \vec{d}_{31} + \rho_{14} \vec{d}_{41} \right) + c.c., \qquad (4)$$

where $N_{ion}^{(res)}$ is the concentration of Ti$^{12+}$ ions, which at the initial moment of time are in state $|1\rangle$; $\rho_{12}$ is the quantum coherence (off-diagonal element of the density matrix) at transition $|1\rangle \leftrightarrow |2\rangle$; $\rho_{13}$ and $\rho_{14}$ are the quantum coherences at transitions $|1\rangle \leftrightarrow |3\rangle, |4\rangle$, respectively; $\vec{d}_{21} = \vec{z}_0 d_z$, $\vec{d}_{31} = \vec{d}_{41} = -i\vec{y}_0 d_y$, $d_z = D/\sqrt{3}$, $d_y = D/\sqrt{6}$, and $D \equiv |\langle 3p^1 S_0 \| D \| 3s^1 P_1 \rangle| \approx 0.41$ a.u. is the reduced dipole moment of the inverted transition. It is worth noting that the dipole moment of transition $|1\rangle \leftrightarrow |2\rangle$ is oriented along the quantization axis $z$, while the dipole moments of transitions $|1\rangle \leftrightarrow |3\rangle, |4\rangle$ lie in the $xy$ plane. As a result, due to the elongation of the active medium along the $x$ axis, transition $|1\rangle \leftrightarrow |2\rangle$ leads to the generation and amplification of the $z$-polarized component of the field $E_z = \vec{z}_0 \cdot \vec{E}$, while transitions $|1\rangle \leftrightarrow |3\rangle, |4\rangle$ lead to that of the $y$-polarized component $E_y = \vec{y}_0 \cdot \vec{E}$. Therefore, in what follows, for brevity, we will call transitions $|1\rangle \leftrightarrow |2\rangle$ and $|1\rangle \leftrightarrow |3\rangle, |4\rangle$ $z$- and $y$-polarized, respectively.

We will describe the evolution of the quantum state of resonant Ti$^{12+}$ ions by a system of equations for the elements of the density matrix $\rho_{ij}$ ($i,j = 1,2,3,4$) of the four-level medium:

$$\begin{aligned}
&\frac{\partial \rho_{11}}{\partial t} + \gamma_{11} \rho_{11} = \frac{i}{\hbar} \sum_{s=1}^{4} (\rho_{s1} \vec{d}_{1s} - \rho_{1s} \vec{d}_{s1}) \vec{E}, \\
&\frac{\partial \rho_{ii}}{\partial t} + \gamma_{ii} \rho_{ii} = A \rho_{11} + \frac{i}{\hbar} \sum_{s=1}^{4} (\rho_{si} \vec{d}_{is} - \rho_{is} \vec{d}_{si}) \vec{E}, \; i \neq 1, \\
&\frac{\partial \rho_{ij}}{\partial t} + [i\omega_{ij}(t,x) + \gamma_{ij}] \rho_{ij} = \frac{i}{\hbar} \sum_{s=1}^{4} (\rho_{sj} \vec{d}_{is} - \rho_{is} \vec{d}_{sj}) \vec{E}, \; i \neq j,
\end{aligned} \qquad (5)$$

where the spatiotemporal dependences of the transition frequencies $|i\rangle \leftrightarrow |j\rangle$ are determined by Eqs. (2), $\vec{d}_{ij}$ is the dipole moment of transition $|i\rangle \leftrightarrow |j\rangle$, $\vec{d}_{ij} \neq 0$ only for $ij = \{12, 21, 13, 31, 14, 41\}$, $A$ is the rate of spontaneous radiative transitions from state $|1\rangle$ to each state $|2\rangle$, $|3\rangle$, and $|4\rangle$, $1/A = 242.5$ ps [33], and $\gamma_{ij}$ are the relaxation rates of the density matrix elements:

$$\gamma_{ii} = \Gamma_{\text{rad}}^{(i)} + w_{\text{ion}}^{(i)}, \; i = 1, 2, 3, 4,$$
$$\gamma_{ij} = (\gamma_{ii} + \gamma_{jj})/2 + \gamma_{\text{Coll}}, \; i \neq j,$$
(6)

where $\Gamma_{\text{rad}}^{(i)}$ is the rate of radiative transitions from state $|i\rangle$ to all lower states: $\Gamma_{\text{rad}}^{(1)} = 50.2$ ps, $\Gamma_{\text{rad}}^{(2)} = \Gamma_{\text{rad}}^{(3)} = \Gamma_{\text{rad}}^{(4)} = 3.3$ ps; $w_{\text{ion}}^{(i)}$ is the rate of tunnel ionization from state $|i\rangle$ under the influence of the IR field, calculated using the Perelomov-Popov-Terentyev formula [34]; $\gamma_{\text{Coll}}$ is the rate of collisional relaxation in plasma, which was calculated from the experimentally measured bandwidth of the gain line of an optically thin medium (see [24]) and is $\gamma_{\text{Coll}}^{-1} = 213$ fs.

Next, we will make a change of variables $t \to \tau = t - x\sqrt{\varepsilon_{\text{pl}}^{(XUV)}}/c$ and look for a solution to the system of equations (3), (4), and (5) using the slowly varying amplitude approximation and the rotating-wave approximation, assuming that

$$\vec{E}(x, \tau) = \frac{1}{2}\left[\vec{z}_0 \tilde{E}_z(x, \tau) + \vec{y}_0 \tilde{E}_y(x, \tau)\right] e^{-i\omega_{XUV}\tau} + \text{c.c.},$$
$$\vec{P}(x, \tau) = \frac{1}{2}\left[\vec{z}_0 \tilde{P}_z(x, \tau) + \vec{y}_0 \tilde{P}_y(x, \tau)\right] e^{-i\omega_{XUV}\tau} + \text{c.c.},$$
$$\rho_{12}(x, \tau) = \tilde{\rho}_{12}(x, \tau) e^{-i\omega_{XUV}\tau}, \; \rho_{13}(x, \tau) = \tilde{\rho}_{13}(x, \tau) e^{-i\omega_{XUV}\tau},$$
$$\rho_{14}(x, \tau) = \tilde{\rho}_{14}(x, \tau) e^{-i\omega_{XUV}\tau},$$
$$\rho_{ij}(x, \tau) = \tilde{\rho}_{ij}(x, \tau), \; ij \neq \{12, 21, 13, 31, 14, 41\},$$
$$\tilde{\rho}_{ij} = \tilde{\rho}_{ji}^*,$$
$$|\omega_{XUV} - \bar{\omega}_z|/\omega_{XUV} \ll 1, \; |\omega_{XUV} - \bar{\omega}_y|/\omega_{XUV} \ll 1,$$
(7)

where $\tilde{E}_{z,y}$ and $\tilde{P}_{z,y}$ are slowly varying complex amplitudes of the polarization components of the generated field and the resonant polarization of the medium, $|\partial \tilde{E}_{z,y}/\partial \tau| \ll \omega_{XUV}|\tilde{E}_{z,y}|$, $|\partial \tilde{E}_{z,y}/\partial x| \ll \omega_{XUV}\sqrt{\varepsilon_{\text{pl}}^{(XUV)}}|\tilde{E}_{z,y}|/c$, $|\partial \tilde{P}_{z,y}/\partial \tau| \ll \omega_{XUV}|\tilde{P}_{z,y}|$, $|\partial \tilde{P}_{z,y}/\partial x| \ll \omega_{XUV}\sqrt{\varepsilon_{\text{pl}}^{(XUV)}}|\tilde{P}_{z,y}|/c$, and $\tilde{\rho}_{ij}$ are slowly varying amplitudes of the elements of the density matrix of the medium, $|\partial \tilde{\rho}_{ij}/\partial \tau| \ll \omega_{XUV}|\tilde{\rho}_{ij}|$, $|\partial \tilde{\rho}_{ij}/\partial x| \ll \omega_{XUV}\sqrt{\varepsilon_{\text{pl}}^{(XUV)}}|\tilde{\rho}_{ij}|/c$. The system of equations obtained in this way is written out in Refs. [24, 27] and, because of its cumbersomeness, is not given here.

The above system of equations must be supplemented with initial and boundary conditions. Since, in contrast to our previous studies [24, 26, 27], the present work investigates the generation of ASE, here the seed radiation at the entrance to the medium is absent, $\tilde{E}_{z,y}(x=0, \tau) = 0$. Since also, in the conditions under consideration, the reflection of XUV radiation from the boundaries of the plasma channel can be neglected, the slowly varying amplitudes of the polarization components of the generated field at the exit from the active plasma of thickness $L$ are determined as $\tilde{E}_{z,y}^{(out)}(\tau) = \tilde{E}_{z,y}(x=L, \tau)$.

In turn, the initial conditions imply that at the time $\tau = 0$, among the considered resonant states, only state $|1\rangle$ is populated, i.e. $\tilde{\rho}_{11}(x, \tau = 0) = 1$, $\tilde{\rho}_{ii}(x, \tau = 0) = 0$ for $i = 2, 3, 4$. To model the ASE source in the active medium, it is assumed that at the initial time, at the inverted transitions $|1\rangle \leftrightarrow |2\rangle, |3\rangle, |4\rangle$, there are random spatial distributions of quantum coherences $\tilde{\rho}_{1i}(x, \tau = 0)$, $i = 2, 3, 4$, while the initial values of the coherences at the remaining transitions are equal to zero. The spatial distributions $\tilde{\rho}_{1i}(x, \tau = 0)$, $i = 2, 3, 4$, are determined as follows [35-37]. The medium is divided into a set of layers $x_{k-1} \leq x < x_k$ of thickness $l_{elem}$, satisfying the condition

$$l_{elem} \leq l_{crit} = \sqrt{\frac{8\pi c}{3\lambda_{XUV} A N_{ion}^{(res)}}}, \tag{8}$$

(in the calculations below it is assumed that $l_{elem} = l_{crit}/2$), in each of which the initial random values of coherences at transitions $|1\rangle \leftrightarrow |2\rangle, |3\rangle, |4\rangle$ are specified as

$$\tilde{\rho}_{12}(x_{k-1} \leq x < x_k, \tau = 0) = \frac{A_{2,k} \exp(i\varphi_{2,k})}{N_{ion}^{(res)} \pi R^2 l_{elem}},$$

$$\tilde{\rho}_{13}(x_{k-1} \leq x < x_k, \tau = 0) = \frac{A_{3,k} \exp(i\varphi_{3,k})}{N_{ion}^{(res)} \pi R^2 l_{elem}}, \tag{9}$$

$$\tilde{\rho}_{14}(x_{k-1} \leq x < x_k, \tau = 0) = \frac{A_{4,k} \exp(i\varphi_{4,k})}{N_{ion}^{(res)} \pi R^2 l_{elem}},$$

where the amplitudes $A_{i,k}$ and phases $\varphi_{i,k}$ ($i = 2, 3, 4$) are independent random variables with the following distribution functions:

$$W(A_{i,k}) = \frac{1}{N_k} \exp\left(-A_{i,k}^2 / N_k\right), \quad 0 \leq A_{i,k}^2 < \infty,$$

$$W(\varphi_{i,k}) = \frac{1}{2\pi}, \quad 0 \leq \varphi_{i,k} < 2\pi, \tag{10}$$

where $N_k = N_{ion}^{(res)} \pi R^2 l_{elem} \tilde{\rho}_{11}(x_{k-1} \leq x < x_k, \tau = 0)$ is the average number of particles that are in the upper state $|1\rangle$ in layer number $k$ at the initial moment of time. In the case of homogeneous active plasma under consideration, $N_k$ does not depend on $k$ and can be expressed, taking into account the initial condition $\tilde{\rho}_{11}(x, \tau = 0) = 1$, as $N_k = N_{ion}^{(res)} \pi R^2 l_{elem}$.

Note that due to the random initial distributions of coherences at resonant transitions, to calculate the physically measurable characteristics of the ASE, such as the spectral power density, $\left|\tilde{S}_{z,y}^{(\omega)}(\omega, L)\right|^2 = \frac{1}{4\pi^2} \left|\int_{-\infty}^{\infty} \tilde{E}_{z,y}(x, \tau) \exp(i\omega\tau) d\tau\right|^2$, and the time dependence of the intensity, $I_{z,y}^{(out)}(\tau) = c\left|\tilde{E}_{z,y}(L, \tau)\right|^2 / (8\pi)$, of the $z$- and $y$-polarized components of the ASE at the output of the plasma channel, it is necessary to perform statistical averaging taking into account the distributions (10). For the analytical solution for $\left|\tilde{S}_{z,y}^{(\omega)}(\omega, L)\right|^2$, which is discussed below, such averaging can be done explicitly. In the numerical solution of the original system of equations, for calculating the ensemble averages $\left|\tilde{S}_{z,y}^{(\omega)}(\omega, L)\right|^2$ and $I_{z,y}^{(out)}(\tau)$, we performed $M$ independent calculations with independently specified random distributions of quantum coherences (9), after which the ensemble averages $\left\langle\left|\tilde{S}_{z,y}^{(\omega)}(\omega, L)\right|^2\right\rangle$ and $\left\langle I_{z,y}^{(out)}(\tau)\right\rangle$ were calculated as

$$\left\langle \left|\tilde{S}_{z,y}^{(\omega)}(\omega,L)\right|^2 \right\rangle = \frac{1}{M}\sum_{m=1}^{M}\left|\tilde{S}_{z,y,m}^{(\omega)}(\omega,L)\right|^2,$$

$$\left\langle I_{z,y}^{(out)}(\tau) \right\rangle = \frac{1}{M}\sum_{m=1}^{M}\left|I_{z,y,m}^{(out)}(\tau)\right|^2, \quad (11)$$

where $\left|\tilde{S}_{z,y,m}^{(\omega)}(\omega,L)\right|^2$ and $I_{z,y,m}^{(out)}(\tau)$ are the spectral power density and the time dependence of the intensity of the *z*- or *y*-polarized ASE component at the output of a plasma channel of thickness *L*, calculated within the *m*-th calculation. Below, we performed averaging over $M = 10$ independent calculations, which, as will be seen below, is sufficient for comparison with the analytical solution.

## III. ANALYTICAL SOLUTION

In order to investigate the main properties of the ASE of optically modulated active medium of a plasma x-ray laser based on neon-like $Ti^{12+}$ ions, we obtained an analytical solution for the spectral power density of the ASE, $\left|\tilde{S}_{z,y}^{(\omega)}(\omega,L)\right|^2$. The solution was obtained under the assumption that the population difference at the transitions $|1\rangle \leftrightarrow |2\rangle, |3\rangle, |4\rangle$ does not change during the ASE generation process. In this case, the *z*- and *y*-polarized ASE components are generated and amplified in the medium independently of each other, and their propagation is described by two independent systems of equations:

$$\begin{cases} \dfrac{\partial \tilde{E}_z}{\partial x} = i\dfrac{4\pi N_{ion}^{(res)} d_z \omega_{XUV}}{c\sqrt{\varepsilon_{pl}^{(XUV)}}}\tilde{\rho}_{12}, \\ \dfrac{\partial \tilde{\rho}_{12}}{\partial \tau} + \left[i(\overline{\omega}_z - \omega_{XUV}) + i\Delta_z \cos(2\Omega\tau + 2\Delta Kx) + \gamma\right]\tilde{\rho}_{12} = -in_{tr}^{(12)}\dfrac{d_z\tilde{E}_z}{2\hbar}, \end{cases} \quad (12)$$

$$\begin{cases} \dfrac{\partial \tilde{E}_y}{\partial x} = \dfrac{4\pi N_{ion}^{(res)} d_y \omega_{XUV}}{c\sqrt{\varepsilon_{pl}^{(XUV)}}}(\tilde{\rho}_{13} + \tilde{\rho}_{14}), \\ \dfrac{\partial \tilde{\rho}_{13}}{\partial \tau} + \left[i(\overline{\omega}_y - \omega_{XUV}) + i\Delta_y \cos(2\Omega\tau + 2\Delta Kx) + \gamma\right]\tilde{\rho}_{13} = n_{tr}^{(13)}\dfrac{d_y\tilde{E}_y}{2\hbar}, \\ \dfrac{\partial \tilde{\rho}_{14}}{\partial \tau} + \left[i(\overline{\omega}_y - \omega_{XUV}) + i\Delta_y \cos(2\Omega\tau + 2\Delta Kx) + \gamma\right]\tilde{\rho}_{14} = n_{tr}^{(14)}\dfrac{d_y\tilde{E}_y}{2\hbar}, \end{cases} \quad (13)$$

where $\Delta K = \Omega(\sqrt{\varepsilon_{pl}^{(XUV)}} - \sqrt{\varepsilon_{pl}^{(IR)}})/c$ is the addition to the wave number of the laser wave, arising from the difference in phase velocities between the ASE at frequency $\omega_{XUV}$ and the modulating field at frequency $\Omega$ due to plasma dispersion; $\gamma = \gamma_{12} \approx \gamma_{13} = \gamma_{14}$, where approximate equality is satisfied under the condition of smallness of the addition to the relaxation rates of coherences due to ionization by the modulating laser field; $n_{tr}^{(1i)} = \tilde{\rho}_{11}(x,\tau=0) - \tilde{\rho}_{ii}(x,\tau=0)$ is the initial population difference at the transition $|1\rangle \leftrightarrow |i\rangle$ ($i = 2,3,4$), which, within the framework of the analytical solution, is assumed to be equal to unity: $n_{tr}^{(1i)} = 1$. In this case, the initial and boundary conditions for the systems of Eqs. (12) and (13) have the form

$$\begin{cases} \tilde{E}_z(x=0,\tau) = 0, \\ \tilde{\rho}_{12}(x,\tau=0) = \rho_{12}^{(a)}(x)e^{i\varphi_{12}(x)}, \end{cases} \quad (14)$$

$$\begin{cases} \tilde{E}_y(x=0,\tau)=0, \\ \tilde{\rho}_{13}(x,\tau=0)=\rho_{13}^{(a)}(x)e^{i\varphi_{13}(x)}, \\ \tilde{\rho}_{14}(x,\tau=0)=\rho_{14}^{(a)}(x)e^{i\varphi_{14}(x)}, \end{cases} \quad (15)$$

where the discrete (in *x*) random initial distributions of coherences (9) are replaced by the corresponding continuous random functions of the coordinate *x*.

As shown in the Appendix, the solutions of the systems of Eqs. (12) and (13) with the initial and boundary conditions (14) and (15), respectively, for the ensemble average spectral power densities of the *z*- and *y*-polarized components of the ASE at the output of a plasma channel of thickness *L* have the form

$$\left\langle \left|\tilde{S}_z^{(\omega)}(\omega,L)\right|^2\right\rangle = S_0^2 \frac{\left|\tilde{G}_z(\omega)\right|^2}{\operatorname{Re}[\tilde{G}_z(\omega)]}\left[e^{2g_0 L \operatorname{Re}[\tilde{G}_z(\omega)]}-1\right],$$

$$\left\langle \left|\tilde{S}_y^{(\omega)}(\omega,L)\right|^2\right\rangle = S_0^2 \frac{\left|\tilde{G}_y(\omega)\right|^2}{\operatorname{Re}[\tilde{G}_y(\omega)]}\left[e^{2g_0 L \operatorname{Re}[\tilde{G}_y(\omega)]}-1\right], \quad (16)$$

where $g_0 = 2\pi\omega_{XUV} N_{\text{ion}}^{(\text{res})} d_z^2 \big/ (\hbar c \gamma \sqrt{\varepsilon_{\text{pl}}^{(XUV)}}) = 4\pi\omega_{XUV} N_{\text{ion}}^{(\text{res})} d_y^2 \big/ (\hbar c \gamma \sqrt{\varepsilon_{\text{pl}}^{(XUV)}})$ is the unperturbed (in the absence of a modulating laser field) gain coefficient of the medium in amplitude, $S_0^2$ is the normalization constant;

$$\tilde{G}_z(\omega)=\sum_{n=-\infty}^{\infty}\frac{J_n^2(P_\Omega^{(z)})}{1+i(\bar{\omega}_z+2n\Omega-\omega-\omega_{XUV})/\gamma},$$

$$\tilde{G}_y(\omega)=\sum_{n=-\infty}^{\infty}\frac{J_n^2(P_\Omega^{(y)})}{1+i(\bar{\omega}_y+2n\Omega-\omega-\omega_{XUV})/\gamma} \quad (17)$$

are the quantities that have the meaning of the normalized to $g_0$ complex gain spectra of the modulated active plasma for the *z*- and *y*-polarized components of the XUV field with the carrier frequency $\omega_{XUV}$, respectively; $P_\Omega^{(z)}=\Delta_z/(2\Omega)$ is the modulation index of the *z*-polarized transition $|1\rangle \leftrightarrow |2\rangle$, and $P_\Omega^{(z)}=\Delta_z/(2\Omega)$ is the modulation index of the *y*-polarized transitions $|1\rangle \leftrightarrow |3\rangle, |4\rangle$. Note that the solution (16), (17) is obtained using two additional conditions. First, it is assumed that the active medium is strongly dispersive for the modulating field, so that $g_0/\Delta K \ll 1$. Second, it is implied that the laser field frequency Ω significantly exceeds the gain line width *γ*: Ω/*γ* >> 1. In what follows, we will consider a modulating field with a wavelength $\Lambda = 2\pi c/\Omega = 1$ µm. This choice is due to the availability of sources of powerful picosecond laser pulses suitable for modulating the active medium of an x-ray laser, with a close wavelength, such as a Ti:Sa laser (0.8 µm), Nd:YAG laser (1.064 µm), and ytterbium lasers (1.03 µm). For the considered parameters of the active plasma of neon-like Ti$^{12+}$ ions and the modulating laser field with a wavelength of 1 µm, both previously mentioned conditions are fulfilled, namely, $g_0/\Delta K \approx 0.02$ (for an unperturbed ASE wavelength of 32.6 nm) and $\Omega/\gamma \approx 400$.

According to the solution (16), (17), the spectra of the *z*- and *y*-polarized components of the generated ASE are determined by the corresponding normalized complex gain spectra $\tilde{G}_z(\omega)$ and $\tilde{G}_y(\omega)$, which properties are governed by the parameters of the modulating laser field, namely, the intensity $I_{IR}$ and the frequency Ω. In the absence of a modulating field, i.e., at $I_{IR}=0$, the modulation indices of the *z*- and *y*-polarized transitions are equal to zero, $P_\Omega^{(z)}=P_\Omega^{(y)}=0$, and the time-averaged frequencies of these transitions coincide and are equal to the unperturbed transition frequency $\bar{\omega}_z = \bar{\omega}_y = \omega_0$, where $\omega_0 \equiv \omega_{12}^{(0)} = \omega_{13}^{(0)} = \omega_{14}^{(0)}$. As a result, the normalized gain spectra

$$\tilde{G}_z(\omega)\big|_{I_{IR}=0} = \tilde{G}_y(\omega)\big|_{I_{IR}=0} = [1+i(\omega_0 - \omega - \omega_{XUV})/\gamma]^{-1} \tag{18}$$

coincide with each other and have the form of a Lorentzian contour centered on the unperturbed transition frequency $\omega_0$ with unit amplitude and full width at half maximum of $2\gamma$ (see black solid line in Fig. 2(a)). As a consequence, the spectra of the $z$- and $y$-polarized components of the ASE also coincide with each other, are single-frequency, and are centered on the frequency $\omega_0$ (see black solid line in Fig. 2(b)):

$$\left\langle \left|\tilde{S}_z^{(\omega)}(\omega,x)\right|^2 \right\rangle\bigg|_{I_{IR}=0} = \left\langle \left|\tilde{S}_y^{(\omega)}(\omega,x)\right|^2 \right\rangle\bigg|_{I_{IR}=0} = S_0^2 \left[ \exp\left(\frac{2g_0 x}{1+i(\omega_0 - \omega - \omega_{XUV})/\gamma}\right) - 1 \right]. \tag{19}$$

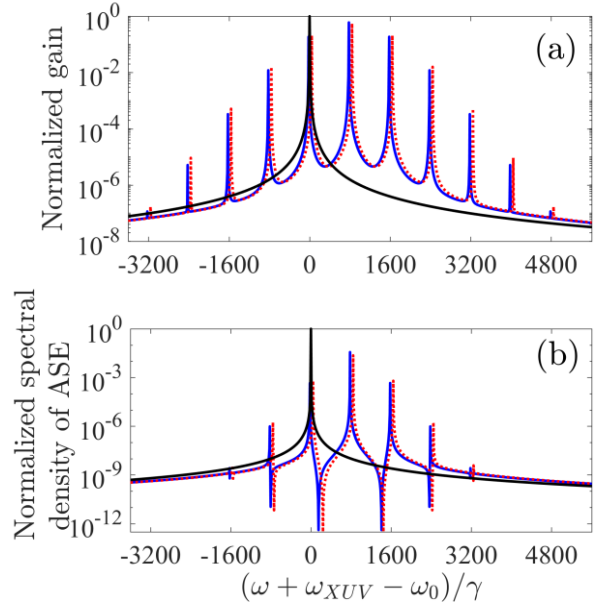

FIG. 2. (a) Normalized gain spectra of the active plasma of neon-like Ti$^{12+}$ ions (17), Re$[\tilde{G}_{z,y}(\omega)]$, and (b) the corresponding normalized spectral power densities of the polarization components of the generated ASE (16), $\left\langle \left|\tilde{S}_{z,y}^{(\omega)}(\omega,x)\right|^2 \right\rangle$, at $g_0 L = 3.5$ ($L = 1$ mm). Blue solid lines in both figures correspond to the gain and ASE spectra of the $z$-polarization component, and the red dotted lines are for the $y$-polarization component. The corresponding parameters of the modulating laser field are $I_{IR} = 2.5 \times 10^{16}$ W/cm$^2$ and $\Lambda = 1$ μm, while $P_\Omega^{(z)} \approx 0.97$ and $P_\Omega^{(y)} \approx 1.05$. Black solid lines in both figures correspond to the gain (18) and ASE (19) spectra in the absence of a modulating field.

Under the action of the modulating laser field, the gain of the medium for both polarization components of the resonant field is redistributed over combination frequencies that are separated from the corresponding time-averaged frequencies of the transitions $\bar{\omega}_z$ and $\bar{\omega}_y$ by an even number of frequencies of the laser field $\Omega$ (Fig. 2(a)). In this case, the gain spectra $\tilde{G}_z(\omega)$ and

$\tilde{G}_y(\omega)$ (17) are sets of Lorentzian contours with amplitudes $J_n^2(P_\Omega^{(z,y)}) < 1$, $\sum_{n=-\infty}^{\infty} J_n^2(P_\Omega^{(z,y)}) = 1$, separated from each other by $2\Omega$. As a result, the spectra of the z- and y-polarized components of the ASE also become multicomponent (Fig. 2(b)). In this case, the gain factors of the spectral components of the ASE are lower than the unperturbed gain factor of the medium $g_0$ by a factor of $J_n^2(P_\Omega^{(z,y)})$, which reduces the efficiency of ASE generation compared to an unmodulated active medium. According to (16), (17), the spectral power densities of the n-th spectral components of the z- and y-polarized ASE at frequencies $\omega + \omega_{XUV} = \bar{\omega}_z + 2n\Omega$ and $\omega + \omega_{XUV} = \bar{\omega}_y + 2n\Omega$, respectively, are determined by

$$S_{n,z}^2 = \left\langle \left| \tilde{S}_z^{(\omega)}(\bar{\omega}_z - \omega_{XUV} + 2n\Omega, L) \right|^2 \right\rangle = S_0^2 J_n^2(P_\Omega^{(z)}) \left[ e^{2g_0 L J_n^2(P_\Omega^{(z)})} - 1 \right],$$

$$S_{n,y}^2 = \left\langle \left| \tilde{S}_y^{(\omega)}(\bar{\omega}_y - \omega_{XUV} + 2n\Omega, L) \right|^2 \right\rangle = S_0^2 J_n^2(P_\Omega^{(y)}) \left[ e^{2g_0 L J_n^2(P_\Omega^{(y)})} - 1 \right].$$

(20)

As follows from (20), the combination spectral components of the ASE are exponentially amplified with increasing medium thickness, $L$. The greatest amplification is for components with maximum gain factors, as a result of which, with increasing thickness of the medium, the number of essentially non-zero spectral components of the ASE decreases, and the spectra of the polarization components of the ASE become narrower than the corresponding gain spectra.

In addition, due to the constant component of the quadratic Stark shift, the centers of the gain spectra and the carrier frequencies of the polarization components of the ASE, $\bar{\omega}_z$ and $\bar{\omega}_y$, are shifted relative to the unperturbed frequency of the inverted transition. Moreover, due to the slight difference between the amplitudes of the Stark modulation of the frequencies of the z- and y-polarized transitions of $Ti^{12+}$ ions, $|\Delta_z|/|\Delta_y| \approx 0.93$, the central frequencies in the gain spectra for the z- and y-polarized field differ by the value $|\bar{\omega}_y - \bar{\omega}_z| = |\Delta_y - \Delta_z|$. Thus, by changing the intensity $I_{IR}$ and the frequency $\Omega$ of the laser field, it is possible to control the frequency position of the combination spectral components of the ASE, as well as to frequency separate its z- and y-polarized components. Note that for neon-like active plasma of $Ti^{12+}$ ions, the shift of the time-averaged frequencies of the z- and y-polarized transitions relative to the unperturbed value, $|\bar{\omega}_{z,y} - \omega_0|$, can be comparable with the frequency of the modulating optical field and exceed the gain line width by 2–3 orders of magnitude. Using the definitions of $\bar{\omega}_z$ and $\bar{\omega}_y$, it can be shown that the shift of the time averaged frequencies of transitions, normalized to the frequency of the modulating field, is equal to the doubled modulation index for transitions of the corresponding polarization: $|\bar{\omega}_{z,y} - \omega_0|/\Omega = 2P_\Omega^{(z,y)}$. For instance, in Fig. 2 at $\Lambda = 1$ μm we have $|\bar{\omega}_z - \omega_0| \approx 780.8\gamma \approx 1.95\Omega$ and $|\bar{\omega}_y - \omega_0| \approx 842.9\gamma \approx 2.11\Omega$. Thus, for a given wavelength of the modulating field $\Lambda$, the frequency shift of the spectra of the polarization components of the ASE increases or decreases proportionally to the intensity of the modulating field $I_{IR}$.

However, with an increase in $I_{IR}$, the ionization rates of the resonant states of $Ti^{12+}$ ions, $w_{ion}^{(i)}$, $i = 1,2,3,4$, also increase, see Fig. 3(a), which leads to the elimination of the population inversion and the disappearance of the amplification of the XUV radiation. As an estimate of the maximum permissible intensity of the modulating laser field, we took the intensity $I_{IR}^{(max)}$ at which the ionization rate of the upper state $|1\rangle$ is equal to 1/10 of the rate of collisional relaxation of the coherence, which determines the half-width of the gain line of the active medium in the absence of modulation, $\gamma_0 \approx \gamma_{\text{Coll}}$: $w_{\text{ion}}^{(1)}(I_{IR}^{(\max)}) = \gamma_{\text{Coll}}/10$. This choice is caused by the fact that the decrease in population inversion due to ionization by the laser field is determined just by the ionization rate of the upper state $|1\rangle$, which significantly exceeds the ionization rates from the lower states $|2\rangle - |4\rangle$, see Fig. 3. According to Fig. 3(a), $I_{IR}^{(\max)} \approx 10^{17}$ W/cm$^2$ and $\left[w_{\text{ion}}^{(1)}(I_{IR}^{(\max)})\right]^{-1} \simeq 2.1$ ps.

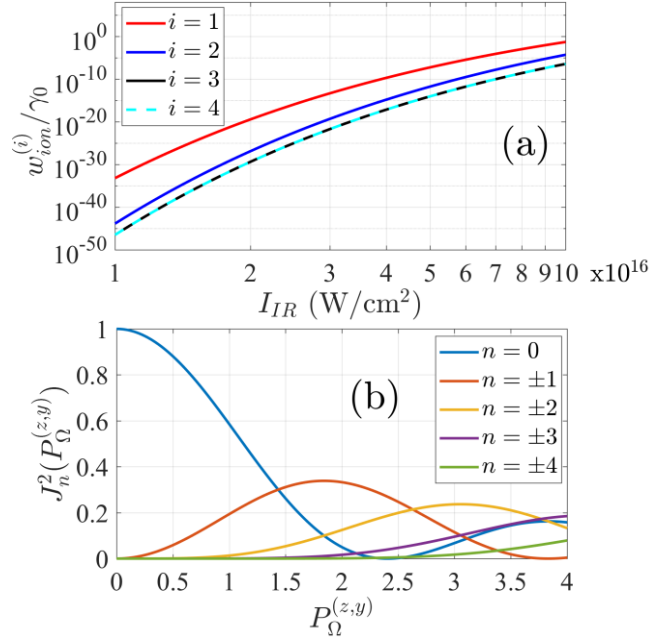

FIG. 3. (a) Dependence of the ionization rate of the state $|i\rangle$, $i = 1,2,3,4$, normalized to the unperturbed value of the coherence relaxation rate at resonance transitions $\gamma_0 = \gamma_{12}(I_{IR} = 0) = \gamma_{13}(I_{IR} = 0) = \gamma_{14}(I_{IR} = 0)$, on the intensity of the modulating field $I_{IR}$ with a wavelength $\Lambda = 1$ μm. (b) Dependence of the square of the Bessel function of the first kind of order $n$ on the modulation index of $z$- or $y$-polarized transitions.

The upper limit of the modulating field intensity also restricts the range of modulation index values that can be achieved in the active plasma under consideration when using a modulating laser field with a given wavelength. For instance, for Ti$^{12+}$ ion plasma and a wavelength of $\Lambda = 1$ μm, the modulation index of the $z$-polarized transition can vary due to a change in the $I_{IR}$ intensity in the range of $0 \leq P_\Omega^{(z)} \leq 3.9$, and the modulation index of the $y$-polarized transitions can vary through $0 \leq P_\Omega^{(y)} \leq 4.2$. As a consequence, the maximum shift (as a whole relative to the unperturbed position) of the $z$-polarized ASE spectrum of the modulated active plasma of Ti$^{12+}$ ions is $|\bar{\omega}_z - \omega_0|/\Omega = 2\max(P_\Omega^{(z)}) = 7.8$, while for the $y$-polarized ASE it is $|\bar{\omega}_y - \omega_0|/\Omega = 2\max(P_\Omega^{(y)}) = 8.4$. In turn, the maximum bandwidth of the ASE is determined by the frequency separation between its outmost spectral components with a substantially nonzero amplitude, which can be estimated as $2\Delta_z = 4P_\Omega^{(z)}\Omega$ and $2\Delta_y = 4P_\Omega^{(y)}\Omega$ for the ASE of $z$- and $y$-polarization, respectively (in an optically thick medium it is a few tens of percent smaller). It should be noted that these quantities are determined exclusively by the intensity of the modulating field and do not depend on its frequency. Thus, for the active plasma of Ti$^{12+}$ ions the maximum bandwidth of the ASE is of the order of 5 petahertz. Further, as can be seen from the analytical solution (20) and Fig. 3(b), the indicated intervals of change in the modulation indices allow two qualitatively different ASE generation regimes to take place.

The first regime corresponds to the generation of quasimonochromatic ASE. This regime takes place at $\max\{P_\Omega^{(z)}, P_\Omega^{(y)}\} = P_\Omega^{(y)} \leq 1$, which corresponds to the condition $J_0^2(P_\Omega^{(z,y)}) \gg J_n^2(P_\Omega^{(z,y)})$, $n \neq 0$. In this case, the gain factors of the central spectral components in the $z$- and $y$-polarizations with $n = 0$ at frequencies $\bar{\omega}_z$ and $\bar{\omega}_y$, respectively, significantly exceed the gain factors of the other spectral components. As a result, at the output of a plasma channel of sufficiently large thickness, the spectrum of the generated ASE in both $z$- and $y$-polarizations will consist of one narrow spectral contour, which is centered at the frequencies $\bar{\omega}_z$ and $\bar{\omega}_y$ for $z$- and $y$-polarizations, respectively. Thus, this range of modulation indices makes it possible to implement the generation of narrowband XUV radiation tunable in frequency over a wide range (up to $\sim 2\Omega$) with the possibility of isolating linear polarization due to the presence of frequency detuning between the $z$- and $y$-polarized components.

The second regime corresponds to the generation of ASE in the form of spectral combs consisting of several sidebands with comparable amplitudes. This regime takes place at $\max\{P_\Omega^{(z)}, P_\Omega^{(y)}\} = P_\Omega^{(y)} > 1$. In this case, at certain values of the modulation indices $P_\Omega^{(z,y)}$, the gain factors of three or more combination spectral components are close to each other. For example, at $P_\Omega^{(z,y)} \approx 1.4$, the gain factors of the spectral components with numbers $n = 0, \pm 1$ (see Fig. 3(b)) are nearly the same and significantly exceed the gain factors of the other spectral components, while at $P_\Omega^{(z,y)} \approx 3.1$ there are 7 such components ($n = 0, \pm 1, \pm 2, \pm 3$). Note that, as in the case of the quasimonochromatic ASE generation regime, the frequency position of the spectral combs can be changed by changing the intensity of the modulating field $I_{IR}$. In this case, the spectral combs of the ASE in the $z$-polarization practically do not overlap with the spectral combs of the ASE in the $y$-polarization.

## IV. NUMERICAL RESULTS

The properties of the ASE of the modulated active plasma of neon-like $Ti^{12+}$ ions described in the previous section were analyzed on the basis of the analytical solution (16), (17), and (20), which does not take into account the influence of the nonlinearity of the medium caused by the change in the population differences at the lasing transitions $|1\rangle \leftrightarrow |2\rangle, |3\rangle, |4\rangle$, as well as the influence of the spectral components on each other [25]. In this section, the influence of these factors on the process of ASE generation by the optically modulated active plasma of $Ti^{12+}$ ions is considered on the basis of the numerical solution of the original system of Maxwell-von Neumann equations (see Section II) for two characteristic cases that correspond to the two generation regimes described above.

Let us begin by considering the case corresponding to the generation regime of the quasimonochromatic ASE. We will assume that the intensity of the modulating field with a wavelength of $\Lambda = 1$ μm is $I_{IR} = 2.5 \times 10^{16}$ W/cm$^2$, which corresponds to the modulation indices $P_\Omega^{(z)} \approx 0.97$ and $P_\Omega^{(y)} \approx 1.05$.

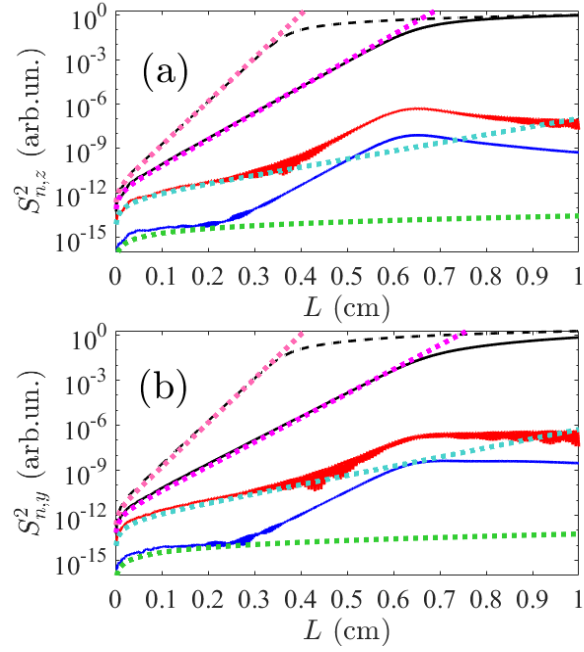

FIG. 4. Dependences of the spectral power density of the ASE components of (a) $z$- and (b) $y$-polarization at frequencies $\bar{\omega}_z \pm 2n\Omega$ and $\bar{\omega}_y \pm 2n\Omega$, respectively, on the thickness $L$ of the optically modulated active plasma of Ti$^{12+}$ ions. Solid lines in each of the figures correspond to the dependences obtained on the basis of the numerical solution of the original system of Maxwell-von Neumann equations (see Section II) at $I_{IR} = 2.5\times10^{16}$ W/cm$^2$ and $\Lambda = 1$ μm ($P_\Omega^{(z)} \approx 0.97$ and $P_\Omega^{(y)} \approx 1.05$) and averaging over $M = 10$ independent realizations, see (11); black color corresponds to the 0-th ASE component in each polarization, i.e. $n = 0$, red color is for $n = 1$, and blue color is for $n = 2$. Dotted lines represent the analytical solution (20) under the same conditions; purple lines correspond to $n = 0$, cyan lines are for $n = 1$, and green lines are for $n = 2$. Dashed black line in both figures corresponds to the spectral power density of the ASE of (a) $z$- and (b) $y$-polarization in the absence of a modulating optical field, i.e., at $I_{IR} = 0$, $P_\Omega^{(z)} = P_\Omega^{(y)} = 0$, calculated via the numerical solution of the original system of Maxwell-von Neumann equations, taking into account averaging over $M = 10$ independent realizations. Dashed pink line corresponds to the analytical solution (20) for $n = 0$ in the absence of a modulating field. The results of the numerical solution in each figure are normalized to the spectral power density of the $z$-polarized ASE component in the absence of a modulating field at $L = 1$ cm. The analytical solution (20) is normalized so that the spectral power density of the $z$-polarized ASE component in the absence of a modulating field at $L = 0.2$ cm coincides with the similar value obtained on the basis of the numerical solution.

In Fig. 4, solid lines show the dependences of the spectral power density of $z$- (Fig. 4(a)) and $y$- (Fig. 4(b)) polarized ASE components at frequencies $\omega + \omega_{XUV} = \bar{\omega}_z \pm 2n\Omega$ and $\omega + \omega_{XUV} = \bar{\omega}_y \pm 2n\Omega$, $n = 0,1,2$, that is $S_{n,z}^2 = \left\langle \left| \tilde{S}_z^{(\omega)}(\bar{\omega}_z - \omega_{XUV} \pm 2n\Omega, L) \right|^2 \right\rangle$ and $S_{n,y}^2 = \left\langle \left| \tilde{S}_y^{(\omega)}(\bar{\omega}_y - \omega_{XUV} \pm 2n\Omega, L) \right|^2 \right\rangle$, see Eq. (11), respectively, on the thickness of the medium $L$. Here, the unperturbed frequency of the resonant transition, $\omega_{XUV} = \omega_0$, is chosen as the carrier frequency of the ASE. For comparison, in these same figures, black dotted lines correspond to the dependences $\left\langle \left| \tilde{S}_z^{(\omega)}(\omega_0 - \omega_{XUV}, L) \right|^2 \right\rangle$ and $\left\langle \left| \tilde{S}_y^{(\omega)}(\omega_0 - \omega_{XUV}, L) \right|^2 \right\rangle$ in the absence of optical modulation of the active medium ($I_{IR} = 0$), and dashed lines correspond to the analytical solution

(20). In this case, the results of the numerical solution are, for convenience, normalized to the value of $\left\langle \left| \tilde{S}_z^{(\omega)}(\omega_0 - \omega_{XUV}, L = 1 \text{ cm}) \right|^2 \right\rangle$, which corresponds to the maximum achievable (within the considered range of $L$) spectral power density of $z$-polarized ASE of the unmodulated active plasma of $Ti^{12+}$ ions. At the same time, the constant in the analytical solution (20) was determined from the condition that the analytical solution for the spectral power density of the $z$-polarized ASE component in the absence of a modulating field at a thickness of $L = 0.2$ cm, where the $z$-polarized ASE is amplified in a linear regime (see Fig. 4(a)), coincides with the similar value calculated on the basis of the numerical solution.

As follows from Fig. 4, there is good qualitative and quantitative agreement between the obtained analytical solution (20) and the numerical solution of the initial system of Maxwell-von Neumann equations at a small thickness of the medium $L$ both in the absence (black dashed and pink dotted lines in Fig. 4) and in the presence (solid and other dotted lines in Fig. 4) of the modulating laser field. At the same time, in the absence of modulation, due to the depletion of the population inversion, starting from $L \approx 0.35$ cm the spectral power density of the ASE reaches saturation.

In the presence of a modulating field, the dependences of the spectral power density of the spectral components of the ASE on $L$ become more complex. If the thickness of the modulated active plasma is small, then due to optical modulation, the spectral components at combination frequencies are independently generated and amplified in it. In this case, due to the comparative smallness of the modulation indices $P_\Omega^{(z,y)}$ and, as a consequence, the smallness of the gain factors of combination frequencies with numbers $n \neq 0$, the spectra of both polarization components of the ASE become narrower with increasing $L$. Each spectrum consists of one intense spectral line at frequencies $\bar{\omega}_z$ and $\bar{\omega}_y$ for the $z$- and $y$-polarized components, respectively, supplemented by several orders of magnitude less intense components at frequencies $\bar{\omega}_z \pm 2n\Omega$ and $\bar{\omega}_y \pm 2n\Omega$ for $n \neq 0$. With further increase in $L$, the increase in the spectral power density of the central spectral component saturates. However, due to the fact that the corresponding gain is lower than the unperturbed value $g_0$ by a factor of $J_0^2(P_\Omega^{(z,y)}) \approx 0.6$, the gain saturation of these components is reached at a larger thickness, $L \approx 0.7$ cm.

At the same time, the dependence of the spectral power density of the sidebands can be divided into three sections. In the first section, the side components are amplified with gain factors proportional to $J_n^2(P_\Omega^{(z,y)})$, $n \neq 0$, in accordance with the analytical solution (20). However, upon reaching a certain thickness of the medium ($L \approx 0.45$ cm for $n = \pm 1$ and $L \approx 0.3$ cm for $n = \pm 2$, see Fig. 4), the spectral power densities of the side spectral components with further growth of $L$ increase at a rate close to the growth rate of the spectral power density of the central ($n = 0$), most intense (for the considered parameters of the modulating field) spectral component, which is determined by its gain factor. This is the second characteristic section of the spatial dependence of the spectral power density of the sidebands of the ASE. In this case, against the background of exponential growth, oscillations arise, the amplitude and period of which decrease with increasing component number $n$ (on the scale of Fig. 4 this corresponds to the thickening of red and blue solid curves). This behavior is explained by the coherent scattering of the central spectral components into the sidebands [19, 21, 25], which is not taken into account in the analytical solution. As shown in [21], when propagating in an optically modulated active medium, narrowband radiation resonant with the central ($n = 0$) gain line is not only amplified with a gain factor proportional to $J_0^2(P_\Omega)$, but also, as a result of the scattering on the modulation wave, generates coherently scattered fields at the frequencies of the side spectral components with $n \neq 0$. The amplitudes and phases of these coherently scattered fields oscillate with an increase in the thickness of the medium $L$ with a characteristic period $\sim [n\Delta K]^{-1}$ due to the difference in the phase velocities of the modulating and resonant fields, which leads to rapid spatial oscillations of the spectral power densities of the combination components of the ASE in Fig. 4.

Thus, in the case considered in this work, the combination spectral components are generated as a result of the combined action of two mechanisms: (a) generation from noise and amplification with gain factors $J_n^2(P_\Omega^{(z,y)})$ and (b) coherent scattering of the most intense component on the modulation wave (which, in turn, is also generated from quantum noise). At small $L$, the dominant mechanism is the generation of sidebands from noise. At large $L$, when the central component has a sufficiently large amplitude, the dominant mechanism becomes coherent scattering. Finally, in the third section, at $L > 0.65$ cm, saturation is observed.

It is also evident from Fig. 4 that the spatial dependences $S_{n,z}^2(L)$ and $S_{n,y}^2(L)$ obtained on the basis of the numerical solution of the initial system of Maxwell-von Neumann equations and averaging over $M = 10$ independent realizations are sufficiently smooth in the sense that they do not exhibit noticeable fluctuations in the region of small thicknesses of the medium $L$, where the random initial phases of coherences in different layers of the active medium $\varphi_{i,k}$ ($i = 2,3,4$), see Eq. (9), play a significant role. This confirms the sufficiency of averaging the physically measured quantities over $M = 10$ independent realizations.

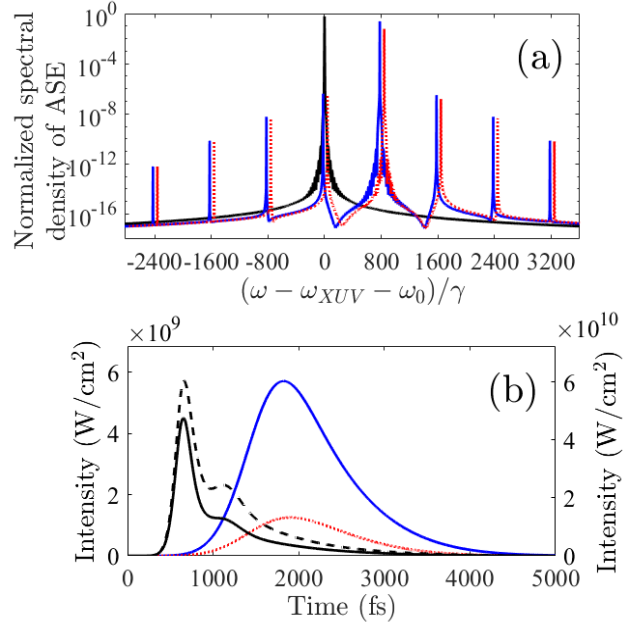

FIG. 5. (a) Normalized spectral power densities and (b) the corresponding time dependences of the intensity of the polarization components of ASE at the output of the active plasma of $Ti^{12+}$ ions with a thickness of $L = 0.7$ cm. Black lines correspond to the absence of the modulating field ($I_{IR} = 0$, $P_\Omega^{(z)} = P_\Omega^{(y)} = 0$); in this case, the spectral contours of ASE for $z$- and $y$-polarizations in (a) coincide (they are not distinguishable on the scale of the figure), while the time dependences in (b) (right vertical axis) differ somewhat due to averaging over a limited ensemble of realizations; black solid line corresponds to the $z$-polarized component of ASE, while black dashed line corresponds to the $y$-polarized component. Blue solid lines in both figures (for (b) see the left

vertical axis) correspond to the $z$-polarized component of ASE in the presence of a modulating laser field with the intensity $I_{IR} = 2.5 \times 10^{16}$ W/cm² and the wavelength $\Lambda = 1$ μm ($P_\Omega^{(z)} \approx 0.97$ and $P_\Omega^{(y)} \approx 1.05$), while red dotted lines correspond to the $y$-polarized component of ASE under the same conditions. In (a) the spectral power is normalized to the peak value for $z$-polarized ASE in the absence of a modulating field. The curves are drawn based on the numerical solution of the Maxwell-von Neumann equations with averaging over $M = 10$ (see (11)) independent realizations.

As a typical example, Fig. 5 shows (a) the normalized spectral power densities and (b) the corresponding time dependences of the intensity of the polarization components of ASE at the output of the active plasma of $Ti^{12+}$ ions with a thickness of $L = 0.7$ cm. Black lines correspond to the absence of the modulating field ($I_{IR} = 0$, $P_\Omega^{(z)} = P_\Omega^{(y)} = 0$), while colored lines correspond to the presence of a modulating laser field with an intensity of $I_{IR} = 2.5 \times 10^{16}$ W/cm² and a wavelength of $\Lambda = 1$ μm ($P_\Omega^{(z)} \approx 0.97$ and $P_\Omega^{(y)} \approx 1.05$). It is evident from Fig. 5(a) that, in accordance with the analytical solution (16), (17), under the action of the modulating laser field, the spectra of the polarization components of the ASE consist of one intense component at frequencies $\bar{\omega}_z$ and $\bar{\omega}_y$, as well as significantly less intense components (approximately by a factor of $10^{-6}$ for the nearest ones, $n = \pm 1$) at frequencies $\bar{\omega}_z \pm 2n\Omega$ and $\bar{\omega}_y \pm 2n\Omega$, $n \neq 0$. As a result, on the scale of Fig. 5(b), the time dependences of the intensity of the polarization components of the ASE have the form of picosecond pulses and do not contain beats caused by the field sidebands. Due to some difference in the values of the modulation indices of the $z$- and $y$-polarized transitions, $P_\Omega^{(z)} < P_\Omega^{(y)}$, the $z$-polarized component of the ASE turns out to be more intense than the $y$-polarized one. In addition, in contrast to the analytical solution (see Fig. 2(b) for comparison), the sidebands turn out to be more intense, and their intensity decreases more slowly with increasing $n$, which is due to the coherent scattering of the central spectral components in both polarizations on the traveling modulation wave.

As mentioned above, at the medium thickness $L = 0.7$ cm and the parameters of the modulating laser field $I_{IR} = 2.5 \times 10^{16}$ W/cm² and $\Lambda = 1$ μm, the gain saturation of the most intense spectral components at the frequencies $\bar{\omega}_z$ and $\bar{\omega}_y$ in both polarizations is reached, see Fig. 4. At the same time, in the absence of the modulating field, the gain saturation of the polarization components of ASE is reached at $L \approx 0.35$ cm, whereas with a further increase in the medium thickness, the spectral power density of the polarization components of ASE remains practically unchanged, while the time dependence of the intensity is shortened due to a rapid depletion in the population inversion at the leading edge of the ASE pulse. Note that in the presence of the modulating field, the same behavior of the time dependences of the pulses of the polarization components of ASE is observed at $L > 0.7$ cm.

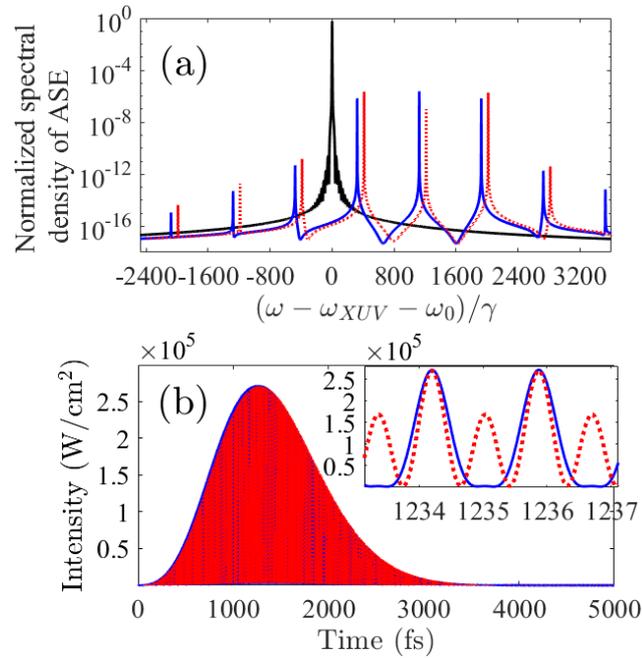

FIG. 6. Same as in Fig. 5, but with a laser field intensity $I_{IR} = 3.6 \times 10^{16}$ W/cm$^2$ ($P_\Omega^{(z)} \approx 1.40$, $P_\Omega^{(y)} \approx 1.52$). The inset in (b) shows the time dependences of the intensities of the polarization components of the ASE on the scale of the period of the modulating field in the vicinity of the maximum of the attosecond beat envelope.

Finally, let us consider the case corresponding to the regime of generating frequency combs in the ASE spectrum. We will choose the parameters of the modulating field so that the gain factors for the 0-th and ±1-st components of the z-polarization ASE are close to each other: $J_0^2(P_\Omega^{(z)}) = J_1^2(P_\Omega^{(z)})$. The first root of this equation is $P_\Omega^{(z)} \approx 1.40$; in this case $P_\Omega^{(y)} \approx 1.52$. For a modulating field with a wavelength of $\Lambda = 1$ μm, such modulation indices are achieved at $I_{IR} = 3.6 \times 10^{16}$ W/cm$^2$. The spatial dependences of the spectral power density of the generated ASE components with such a choice of modulating field parameters are similar to those considered earlier. We only note that due to the larger value of the modulation indices and, as a consequence, smaller values of the gain factors, at the thicknesses under consideration of $L \leq 1$ cm, gain saturation for the spectral ASE components is not reached.

Figure 6 shows the corresponding spectra and time dependences of the polarization components of the ASE at the output of the modulated active plasma of Ti$^{12+}$ ions with a thickness of $L = 0.7$ cm. It is evident that, unlike the case shown in Fig. 5, the spectra of z- and y-polarized ASE consist of three spectral components with $n = 0, \pm 1$, of comparable amplitudes with the total bandwidth of 1.2 petahertz, which form subfemtosecond beats in the time domain. In this case, due to the proximity of the gain factors of the spectral components of the z- and y-polarization, the intensities of the polarization components are close to each other. Since $J_0^2(1.40) \approx J_1^2(1.40)$, the gain factors for the central and side spectral components of the z-polarization ASE practically coincide, and the beats in the time dependence of its intensity are caused by the interference of three components close in amplitude. As follows from Fig. 6(b), a sequence of subfemtosecond pulses is formed in the z-polarization, which indicates the phase matching of the generated spectral components. A similar picture is observed for the y-polarization ASE. However, since $J_0^2(1.52) < J_1^2(1.52)$, the beats in the time dependence of its intensity actually correspond to the interference of the side components at frequencies $\bar{\omega}_y \pm 2\Omega$, which are amplified significantly more efficiently than the central component at frequency $\bar{\omega}_y$. It is important to note that the relative amplitudes and phases of the spectral components of the ASE, as well as the shape of the subfemtosecond pulses (beats) they form in the time dependence of the radiation intensity, are deterministic (do not depend on random initial conditions (9), (10)) and do not change from one realization to another, i.e. for each individual realization, only the absolute values of the amplitudes of the spectral components and the beat intensities differ.

## V. CONCLUSION

In this work, using the example of active plasma of neon-like $Ti^{12+}$ ions, the possibility of generating spectral combs of petahertz bandwidth from amplified spontaneous emission (ASE) of a collisional plasma-based x-ray laser under irradiation of its active medium by a linearly polarized laser field of the near-IR range is investigated. An analytical solution for the spectral power density of the ASE of such an active medium is obtained. Based on this analytical solution and the numerical solution of the Maxwell-von Neumann equations, the spatial, spectral, temporal, and polarization properties of the ASE are analyzed. It is shown that the action of the strong modulating field of linear polarization leads to anisotropy of the active medium of the x-ray laser and a difference in the properties (amplitude and frequency discrimination) of the polarization components of the ASE parallel and orthogonal to the modulating field. In this case, with an increase in the intensity of the modulating field, the thickness of the active medium required to saturate the growth of the ASE amplitude increases. As long as the depth of frequency modulation of the inverted transitions as a result of the quadratic Stark effect does not exceed the doubled frequency of the modulating field, the ASE remains quasimonochromatic, but its carrier frequency changes (increases) relative to its unperturbed value within the doubled modulation frequency (which can exceed the gain spectrum width of the x-ray laser in the absence of modulation by 2–3 orders of magnitude) due to the constant component of the Stark shift. At the same time, in a more intense modulating field, the ASE spectrum is enriched with combination components spaced from the ASE central frequency by an even number of modulation frequencies, which allows generating petahertz-wide spectral combs of XUV radiation. In this case, the interference of the spectral components of ASE leads to beats in the time dependence of its intensity on the scale of the modulating field cycle. These beats have a deterministic form, which is determined exclusively by the characteristics of the modulating field and does not depend on fluctuations of the initial values of coherences on the inverted transitions of the active medium, which are the source of ASE. It is shown that under certain conditions, the above-mentioned beats take the form of a sequence of subfemtosecond pulses.

The proposed scheme can be considered as the simplest case for experimental implementation, implying modulation of the active medium of the plasma-based x-ray laser by the IR field, which can be a kind of prologue to experiments on amplification of the XUV/x-ray radiation of the seed and its spectral broadening. On the other hand, the results of this work can be of independent practical interest for the creation of a source of frequency-tunable quasimonochromatic radiation of the XUV/x-ray range or petahertz frequency combs and subfemtosecond pulse trains.

To implement this approach, the neon-like ions other than $Ti^{12+}$ can be used [30], as well as, due to the similarity of the energy structure, the nickel-like ions, in particular, those that allow the generation of soft x-ray radiation with a wavelength of less than 10 nm [7].

## ACKNOWLEDGMENT

The work was supported by the Foundation for the Development of Theoretical Physics and Mathematics "BASIS", Grant No. 24-1-2-43-1.

## APPENDIX

Let us begin by considering the generation of the $z$-polarized component of the ASE, which is described by the system of equations (12) with initial and boundary conditions (14). To solve the equation with respect to $\tilde{\rho}_{12}$, we use the method of generalized functions [38]. According to this method, the sought function is replaced by a function that coincides with it for $\tau > 0$ and is equal to zero $\tau < 0$. For this new function, the initial equation turns out to be valid, in the right-hand side of which a singular addition appears, equal to the product of the Dirac $\delta$-function and the initial value $\tilde{\rho}_{12}$. Thus, the second equation of the system (12) takes the form

$$\frac{\partial \tilde{\rho}_{12}}{\partial \tau} + \left\{ i(\bar{\omega}_z - \omega_{XUV}) + i\Delta_z \cos(2\Omega\tau + 2\Delta Kx) + \gamma \right\} \tilde{\rho}_{12} = -i \frac{d_z \tilde{E}_z}{2\hbar} + \delta(\tau) \rho_{12}^{(a)}(x) e^{i\varphi_{12}(x)}. \tag{A1}$$

We will look for its solution in the form

$$\tilde{\rho}_{12}(x,\tau) = \hat{\rho}_{12}(x,\tau) \exp\left[ -iP_\Omega^{(z)} \sin(2\Omega\tau + 2\Delta Kx) \right], \tag{A2}$$

where $P_\Omega^{(z)} = \Delta_z/(2\Omega)$ is the modulation index of the $z$-polarized transition $|1\rangle \leftrightarrow |2\rangle$, and the function $\hat{\rho}_{12}(x,\tau)$ satisfies the equation

$$\frac{\partial \hat{\rho}_{12}}{\partial \tau} + \left\{ i(\bar{\omega}_z - \omega_{XUV}) + \gamma \right\} \hat{\rho}_{12} = \left[ -i \frac{d_z \tilde{E}_z}{2\hbar} + \delta(\tau) \rho_{12}^{(a)}(x) e^{i\varphi_{12}(x)} \right] e^{iP_\Omega^{(z)} \sin(2\Omega\tau + 2\Delta Kx)}. \tag{A3}$$

Next, we represent $\hat{\rho}_{12}$ and $\tilde{E}_z$ as the Fourier integrals:

$$\hat{\rho}_{12}(x,\tau) = \int_{-\infty}^{\infty} \hat{\rho}_{12}^{(\omega)}(x,\omega) e^{-i\omega\tau} d\omega,$$

$$\tilde{E}_z(x,\tau) = \int_{-\infty}^{\infty} \tilde{S}_z^{(\omega)}(x,\omega) e^{-i\omega\tau} d\omega, \tag{A4}$$

where $\tilde{S}_z^{(\omega)}(x,\omega)$ is the spectral amplitude of the $z$-polarized component of the ASE, and $\tilde{S}_z^{(\omega)}(x=0,\omega) = 0$. Substituting (A4) into (A3) and taking into account that $\frac{1}{2\pi} \int_{-\infty}^{\infty} \delta(\tau) e^{iP_\Omega^{(z)} \sin(2\Omega\tau + 2\Delta Kx)} e^{i\omega\tau} d\tau = \frac{1}{2\pi} e^{iP_\Omega^{(z)} \sin(2\Delta Kx)}$, we obtain

$$\hat{\rho}_{12}^{(\omega)}(x,\omega) = -i \frac{d_z}{2\hbar\gamma} \sum_{n=-\infty}^{\infty} J_n(P_\Omega^{(z)}) e^{i2n\Delta Kx} \frac{\tilde{S}_z^{(\omega)}(x,\omega + 2n\Omega)}{1 + i(\bar{\omega}_z - \omega_{XUV} - \omega)/\gamma} +$$

$$+ \frac{1}{2\pi\gamma} \frac{\rho_{12}^{(a)}(x) e^{i\varphi_{12}(x)}}{1 + i(\bar{\omega}_z - \omega_{XUV} - \omega)/\gamma} \sum_{n=-\infty}^{\infty} J_n(P_\Omega^{(z)}) e^{i2n\Delta Kx}, \tag{A5}$$

where $J_n(x)$ is the Bessel function of the first kind of order $n$, and the equality $\exp[iP\sin(x)] = \sum_{n=-\infty}^{\infty} J_n(P) e^{inx}$ was used. Next, calculating the direct Fourier transform of the left-hand and right-hand parts of (A2), we obtain the following relationship between $\tilde{\rho}_{12}^{(\omega)}(x,\omega) = \frac{1}{2\pi}\int_{-\infty}^{\infty}\tilde{\rho}_{12}(x,\tau)e^{i\omega\tau}d\tau$ and $\hat{\rho}_{12}^{(\omega)}(x,\omega)$:

$$\tilde{\rho}_{12}^{(\omega)}(x,\omega) = \sum_{k=-\infty}^{\infty} J_k(P_\Omega^{(z)})e^{-i2k\Delta Kx}\hat{\rho}_{12}^{(\omega)}(x,\omega - 2k\Omega). \quad (A6)$$

Substituting (A4), (A5), and (A6) into the first equation of system (12), we obtain an equation for the spectral amplitude $\tilde{S}_z^{(\omega)}(x,\omega)$:

$$\frac{\partial \tilde{S}_z^{(\omega)}}{\partial x} = g_0 \tilde{G}_z(\omega)\tilde{S}_z^{(\omega)} + \frac{i\hbar}{\pi d_z} g_0 \rho_{12}^{(a)}(x)e^{i\varphi_{12}(x)} \sum_{n,k=-\infty}^{\infty} \frac{J_n(P_\Omega^{(z)})J_k(P_\Omega^{(z)})e^{i2(n-k)\Delta Kx}}{1+i(\bar{\omega}_z + 2k\Omega - \omega - \omega_{XUV})/\gamma}, \quad (A7)$$

where $g_0 = 2\pi\omega_{XUV}N_{\text{ion}}^{(\text{res})}d_z^2/(\hbar c\gamma\sqrt{\varepsilon_{\text{pl}}^{(XUV)}})$ is the unperturbed (in the absence of a modulating laser field) gain of the medium, and $\tilde{G}_z(\omega)$ is the gain of the z-polarization field at frequency $\omega$, normalized to $g_0$, determined by

$$\tilde{G}_z(\omega) = \sum_{n=-\infty}^{\infty} \frac{J_n^2(P_\Omega^{(z)})}{1+i(\bar{\omega}_z + 2n\Omega - \omega - \omega_{XUV})/\gamma}, \quad (A8)$$

Note that Eq. (A7) is obtained under the assumption of strong plasma dispersion at the frequency of the modulating laser field, namely, the fulfillment of the inequality

$$g_0/\Delta K \ll 1. \quad (A9)$$

In this case, the spectral components of the ASE corresponding to different $n$ are amplified independently of each other [21, 24-27]. For the considered concentration of free electrons in the plasma $N_e = 5\times 10^{19}$ cm$^{-3}$, the wavelength of the laser field $\Lambda = 2\pi c/\Omega = 1$ μm, and the unperturbed wavelength of the ASE 32.6 nm, we have $g_0/\Delta K \approx 0.02$.

The solution to Eq. (A7) is

$$\tilde{S}_z^{(\omega)}(\omega, x) = \frac{i\hbar g_0}{\pi d_z} \sum_{n,k=-\infty}^{\infty} \frac{J_n(P_\Omega^{(z)})J_k(P_\Omega^{(z)})e^{g_0\tilde{G}_z(\omega)x}}{1+i(\bar{\omega}_z + 2k\Omega - \omega - \omega_{XUV})/\gamma} \int_0^x \rho_{12}^{(a)}(x')e^{i\varphi_{12}(x')}e^{[i2(n-k)\Delta K - g_0\tilde{G}_z(\omega)]x'}dx'. \quad (A10)$$

Expression (A10) represents the complex spectrum of z-polarized ASE generated by the modulated active plasma of neon-like ions. However, in the experiment the measured quantity is the ensemble-averaged spectral power density, determined by

$$\left\langle \left|\tilde{S}_z^{(\omega)}(\omega,x)\right|^2 \right\rangle = \frac{\hbar^2 g_0^2 e^{2g_0 x \text{Re}[\tilde{G}_z(\omega)]}}{\pi^2 d_z^2} \times$$

$$\times \sum_{n_1,k_1,n_2,k_2=-\infty}^{\infty} \frac{J_{n_1}(P_\Omega^{(z)})J_{k_1}(P_\Omega^{(z)})J_{n_2}(P_\Omega^{(z)})J_{k_2}(P_\Omega^{(z)})}{[1+i(\bar{\omega}_z + 2k_1\Omega - \omega - \omega_{XUV})/\gamma][1-i(\bar{\omega}_z + 2k_2\Omega - \omega - \omega_{XUV})/\gamma]} \times \quad (A11)$$

$$\times \int_0^x dx' \int_0^x dx'' \left\langle \rho_{12}^{(a)}(x')\rho_{12}^{(a)}(x'')\right\rangle \left\langle e^{i\varphi_{12}(x')-i\varphi_{12}(x'')}\right\rangle e^{[i2(n_1-k_1)\Delta K - g_0\tilde{G}_z(\omega)]x' + [-i2(n_2-k_2)\Delta K - g_0\tilde{G}_z^*(\omega)]x''}.$$

Since at different points of the active medium the phases $\varphi_{12}$ are statistically independent random quantities with a uniform probability distribution, then $\left\langle e^{i\varphi_{12}(x')-i\varphi_{12}(x'')}\right\rangle = \delta(x' - x'')$. This, as well as the fact that the frequency $\Omega$ of the IR laser field significantly exceeds $\gamma$, $\Omega/\gamma \gg 1$ (at $\Lambda = 1$ μm $\Omega/\gamma \approx 400$), allows us to simplify expression (A11):

$$\left\langle \left| \tilde{S}_z^{(\omega)}(\omega,x) \right|^2 \right\rangle = \frac{\hbar^2 g_0 C_z}{2\pi^2 d_z^2} \frac{\left| \tilde{G}_z(\omega) \right|^2}{\text{Re}[\tilde{G}_z(\omega)]} \left[ e^{2g_0 x \text{Re}[\tilde{G}_z(\omega)]} - 1 \right], \tag{A12}$$

where inequality (A9) is also taken into account, and $C_z = \left\langle \rho_{12}^{(a)}(x)^2 \right\rangle$ is a constant, which, according to (9) and (10), is determined by the average number of particles in an elementary layer (in this case, by the thickness $dx$) in the state $|1\rangle$ and is inversely proportional to it.

Similarly, the system of Eqs. (13) and (15) can be solved, and an expression for the ensemble-averaged spectral power density of the $y$-polarized component of the ASE can be obtained:

$$\left\langle \left| \tilde{S}_y^{(\omega)}(\omega,x) \right|^2 \right\rangle = \frac{\hbar^2 g_0 C_y}{4\pi^2 d_y^2} \frac{\left| \tilde{G}_y(\omega) \right|^2}{\text{Re}[\tilde{G}_y(\omega)]} \left[ e^{2g_0 x \text{Re}[\tilde{G}_y(\omega)]} - 1 \right], \tag{A13}$$

where $C_y = \left\langle \rho_{13}^{(a)}(x)^2 \right\rangle = \left\langle \rho_{14}^{(a)}(x)^2 \right\rangle = C_z$ since the distribution functions for the initial values of coherence amplitudes are the same (see Eqs. (9), (10)), and $\tilde{G}_y(\omega)$ is the normalized (to $g_0$) gain spectrum of the $y$-polarized component of the XUV field:

$$\tilde{G}_y(\omega) = \sum_{n=-\infty}^{\infty} \frac{J_n^2(P_\Omega^{(y)})}{1 + i(\bar{\omega}_y + 2n\Omega - \omega - \omega_{XUV})/\gamma}, \tag{A14}$$

$P_\Omega^{(y)} = \Delta_y/(2\Omega)$ is the modulation index of $y$-polarized transitions $|1\rangle \leftrightarrow |3\rangle, |4\rangle$.

Thus, taking into account that $d_z^2 = 2d_y^2$, the normalization constants in expressions (A12) and (A13) are the same, and these expressions can be rewritten in their final form:

$$\left\langle \left| \tilde{S}_z^{(\omega)}(\omega,x) \right|^2 \right\rangle = S_0^2 \frac{\left| \tilde{G}_z(\omega) \right|^2}{\text{Re}[\tilde{G}_z(\omega)]} \left[ e^{2g_0 x \text{Re}[\tilde{G}_z(\omega)]} - 1 \right],$$

$$\left\langle \left| \tilde{S}_y^{(\omega)}(\omega,x) \right|^2 \right\rangle = S_0^2 \frac{\left| \tilde{G}_y(\omega) \right|^2}{\text{Re}[\tilde{G}_y(\omega)]} \left[ e^{2g_0 x \text{Re}[\tilde{G}_y(\omega)]} - 1 \right], \tag{A15}$$

where $S_0^2 = \hbar^2 g_0 C_z/(2\pi^2 d_z^2) = \hbar^2 g_0 C_y/(4\pi^2 d_y^2)$. Solution (A8), (A14), and (A15) coincides with expressions (16) and (17) in the main text.